\documentclass{emulateapj}
\usepackage{txfonts}
\usepackage{natbib,graphicx,amssymb}
\usepackage[english]{babel}
\usepackage{mathrsfs}
\usepackage{url}
\usepackage{float}
\usepackage{times}

\newcommand{\ha}{\hbox{H$\alpha$}}

\newcommand{\oii}{\hbox{[O\,{\sc ii}]}}
\begin{document}

\title
{The quenching timescale and quenching rate of galaxies}

\author{Jianhui Lian\altaffilmark{1,2}, Renbin Yan\altaffilmark{2}, Kai Zhang\altaffilmark{2}, Xu Kong\altaffilmark{1}}
\email{lianjianhui@uky.edu (J. Lian); renbin@pa.uky.edu (R. Yan); xkong@ustc.edu.cn (X. Kong)}
\altaffiltext{1}{ CAS Key Laboratory for Research in Galaxies and Cosmology, Department of Astronomy,
University of Science and Technology
of China, Hefei, Anhui 230026, China}
\altaffiltext{2}{Department of Physics and Astronomy, University of Kentucky, Lexington, Kentucky 40506, U.S.}

\begin{abstract}
The average star formation rate (SFR) in galaxies has been declining since redshift of 2.
A fraction of galaxies quench and become quiescent. 
We constrain two key properties of the quenching process: the quenching time scale and the quenching rate among galaxies.
We achieve this by analyzing the galaxy number density profile in NUV$-u$ color space and the distribution in 
NUV$-u$ v.s. $u-i$ color-color diagram with a simple toy-model framework.
We focus on galaxies in three mass bins between $10^{10}$ and $10^{10.6} {\rm M_{\odot}}$.
In the NUV$-u$ v.s. $u-i$ color-color diagram, the red $u-i$ galaxies exhibit a different slope from the slope traced by the 
star-forming galaxies. This angled distribution and the number density profile of galaxies in NUV$-u$ space strongly suggest
that the decline of the SFR in galaxies has to accelerate before they turn quiescent. We model this color-color distribution with a 
two-phase exponential decline star formation history. The models with an e-folding time in the second phase (the quenching 
phase) of 0.5 Gyr best fit the data. 
We further use the NUV$-u$ number density profile to constrain the quenching rate among star-forming galaxies as a function of 
mass. 
Adopting an e-folding time of 0.5 Gyr in the second phase (or the quenching phase), we found the quenching rate to be 
19\%/Gyr, 25\%/Gyr and 33\%/Gyr for the three mass bins. These {are} upper limits of quenching rate 
as the transition zone could also be populated by rejuvenated red-sequence galaxies.
\end{abstract}

\keywords{galaxies: evolution -- galaxies: photometry -- galaxies: star formation.}

\section{Introduction}

In many large-scale surveys, the bimodal distribution of galaxy population   
has been found in color-magnitude and color-mass diagrams (e.g., \citealt{strateva2001,baldry2004}).
In these diagrams, galaxies located between the blue and red populations
were often called `green valley galaxies'. 
In addition to colors, many other properties of the green valley galaxies also exhibit intermediate value 
between the two main populations, such as spectral indices 
\citep{kauffmann2003} and morphological parameters \citep{driver2006,pan2013}.
Therefore, the green valley galaxies were considered to represent the transition population 
from the blue star-forming to red-sequence galaxies \citep{bell2004,faber2007,mendez2011,goncalves2012}.
However, some studies reported that 
dusty star-forming galaxies may also exhibit intermediate colors and potentially contaminate the transition 
population \citep{brammer2009,salim2009}. 
{In addition, by differentiating the morphology of green valley galaxies based on Galaxy Zoo project,
\citet{scha2014} and \citet{smethurst2015} argued multiple evolution pathways of galaxies through the green valley zone,
with early-type galaxies quenching and late-type galaxies stalling.
That conclusion could be sensitive to the exact definition (choice of color and range) of the green 
valley, and is made under the assumption of unchanging morphology.
It is also possible that some of the green valley galaxies may come from the red sequence due to 
rejuvenated star formation. But we expect this fraction to be small \citep{fang2012} 
and we expect galaxies going from
green valley to the blue cloud to be even rarer, although not impossible.}

Studies about the evolution of luminosity (or stellar mass) function of different galaxy types with redshift support 
this transition scenario. 
From $z$=1 to 0, the number density of red galaxies have increased significantly by a 
factor of 2 or more \citep{blanton2006,brown2007,faber2007} while 
that of blue galaxies decreased slightly or barely changed \citep{faber2007,ramos2011,moustakas2013}. 
This differential evolution of luminosity function with galaxy types
suggests that many blue galaxies have ceased their star formation and evolved into red sequence during this time period.
The rarity of green valley population implies that the transition timescale must be 
short \citep{faber2007,martin2007,balogh2011}.
For example, 
post starburst galaxies were considered to be the one of the possible candidates of galaxies that are transiting from 
blue to red population {\citep{yange2008, wong2012}}. Their spectra show strong Balmer absorption lines 
but negligible \oii\ or \ha\ emission lines, suggesting a violent shutdown of star formation in the recent past with 
a large amount of A stars still present. However, the small fraction of post-starburst galaxies means they can only 
account for a small fraction of the increase in the number density profile of red galaxies. The majority 
of galaxies may have ceased their SF more smoothly. 
Before turning quiescent,
the star formation rate (SFR) in galaxies have been gradually declining since redshift of 2 \citep{madau2014}.
This gradual decline of SFR introduces the cosmic evolution of main-sequence relation (i.e. mass-SFR relation 
\citealt{daddi2007,elbaz2007,noeske2007a}).

In this work, concerning the quenching process, the problems we would like to address are:
(1) how fast do galaxy SFR have to decline to turn quiescent? Does it have to be much faster than the long term decline?
(2) What is the quenching rate (the fraction of quenching in a given period) among galaxies?
{For the evolution speed during quenching, 
\citet{wetzel2012,wetzel2013} examined the star formation histories of local satellite galaxies and 
obtained a short quenching e-folding time ($<0.8$ Gyr) using a cosmological $N$-body simulation.
Based on EAGLE cosmological hydrodynamical simulation, \citet{trayford2016}
found the time scale to cross the green valley to be less than 2 Gyrs.}
Recently, based on the {\em GALEX} and SDSS data, \citet{scha2014} investigated the green valley galaxies in 
NUV$-u$ v.s. $u-i$ color-color diagram and 
found that early-type green valley galaxies 
evolve much faster than the late-type ones with a shorter e-folding time.
However the analysis based on morphology-separated populations may introduce bias in the result because the galaxy morphology may change 
significantly along with the transition from blue to red galaxies.
In this work, we revisit the UV$-$optical color-color diagrams to answer how fast the SFR has to decline
when a galaxy evolves through the
transition (or green valley) zone (\textsection3).

In terms of the quenching rate, \citet{moustakas2013} obtained it as a function of stellar mass and redshift 
by comparing the stellar mass function of star-forming 
and passive galaxies at different redshift.
{They measured the stellar mass function from $z$=0-1 based on large galaxy sample from
PRism MUlti-object Survey for intermediate-$z$ and SDSS for low-$z$.
However, the quenching rate determined in this way rely on how well the stellar mass function is measured which 
may be limited by sample variance and observation accurancy at high redshift. Also, systematical errors may be hard to avoid
, although they did really careful work to addresss it,
when perform comparison between the two large surveys. Using simply divided star-forming and passive populations,
the information of transition population in between is neglected.}
{In this work, we go beyond simple SFR/color cuts to examine the nature of full color distribution.}
Throughout this paper, we adopted the cosmological parameters with $H_0=70\, {\rm km s^{-1} Mpc}^{-1}$, $\Omega_{\Lambda}=0.73$ 
and $\Omega_{\rm m}=0.27$. All magnitudes in this paper are given in the AB photometric system.

\section{Data}
While the optical light of a galaxy is dominated by intermediate- and low-mass stars,
its ultraviolet light typically traces the emission by young massive stars. Therefore the 
UV observations provide a good probe of the recent SFR. 
Combining the UV and optical observations, we are able to investigate the rate at which the 
specific star formation rates (sSFR) decrease
when galaxies evolve through the transition zone.
{\sl Galaxy Evolution Explorer} ({\em GALEX}; \citealt{martin2005}) mapped {more than half of} the sky at FUV and NUV,
which enables us 
to study the SFR decline in green valley galaxies with a large and complete sample.
The SDSS imaging survey mapped $14,000$ square degrees of the sky
in five broad optical bands: $u$, $g$, $r$, $i$ and $z$.
The depth of the imaging survey reaches $21.3$ mag at $i$ band.
The SDSS spectroscopic survey obtained optical spectra for
galaxies brighter than 17.7 mag in $r$ band with a sky coverage of $\sim$ 8000 square degrees. 
Utilizing the {\em GALEX} and SDSS imaging surveys, 
{based on a consistent and unified analysis of galaxies,}
\citet{blanton2011} created the NASA-Sloan Atlas (NSA) which includes both the photometry catalog and multiband images from UV 
to near-IR.
In the photometry catalog, \citet{blanton2011} applied a better treatment of sky subtraction to 
obtain more accurate photometry for nearby spatially-resolved galaxies.
{For the UV part, all available GALEX observations was used, including 
the Nearby Galaxy Survey (NGS), Deep (DIS), Medium (MIS), and All Sky Imaging
Surveys (AIS).}
It should be noted that the performance of photometry and astrometry is relative poor at the edge of field of view (FOV)
of {\sl GALEX} \citep{morrissey2007,drinkwater2010}, which could potentially affect the UV
photometry and astrometric matching with optical data of our sample. However, given a circle FOV of {\sl GALEX} with a diameter of $1.25^{\circ}$, 
the area of edge (annulus with $r>0.575^{\circ}$) only covers 15\% of the whole FOV. Therefore, we expect 
this edge effect can affects at most 15\% of our sample and it should not change the results significantly.

To select our sample, we require the galaxies to have:
\begin{enumerate}
\item spectra classified as `GALAXY'
\item 0.02$<z<$0.05
\item stellar mass higher than $10^9\ {\rm M_{\odot}}$
\item axis ratio greater than 0.7
\end{enumerate}
The upper redshift limit is set to ensure that the sample is complete in mass. 
The lower limit is set to avoid very nearby galaxies that are large in the sky, for which 
the sky subtraction and deblending is difficult.
The high axis ratio selects face-on galaxies to minimize the effect caused by internal dust extinction.
Our final sample consists of 25,074 objects.
We use the elliptical Petrosian magnitude from UV to optical in the NSA catalog \footnote{http://www.nsatlas.org/data}. 
The $K$-corrections and Galactic dust extinctions are also taken from the NSA catalog.  
We do not correct UV and optical fluxes for internal dust extinction because such
corrections carry large systematical uncertainties and tend to be unreliable.
To account for dust without introducing additional uncertainties to the data,
we apply dust extinction to stellar population synthesis models to qualitatively describe the extinction suffered
by the observed colors.
The measurements of stellar mass were taken from the MPA-JHU catalog \citep{brinchmann2004}.

\section{How fast do galaxies evolve through transition zone}

\subsection{stellar population synthesis model}
It is still under debate whether galaxies experience an accelerated SFR decline before turning 
quiescent. In this work, we investigate both one-phase and two-phase {SFHs} based on stellar population 
synthesis models. For simplicity, we assume galaxy SFR decline exponentially in each evolution phase.
Besides a single exponential decline model as a special case, we 
generally assume galaxies experience a two-phase evolution to evolve into red sequence.
In the first star-forming stage, the SFR declines slowly with a long e-folding time.
While at some point, the SFR begin to drop rapidly with a shorter e-folding time
which we refer to as `quenching' stage. 
In this paper, we differentiate the term describing galaxies going through a particular evolution stage in the model from the 
term describing galaxies satisfying an observational definition (e.g. a color cut) since there may not be a strict correspondence
between the two. In particular, we refer to galaxies going through the second evolution phase as `quenched galaxies'.
And we refer to galaxies defined by having intermediate colors as `transition galaxies' or `green-valley galaxies'.

\begin{figure*}
\centering
\includegraphics[width=15cm]{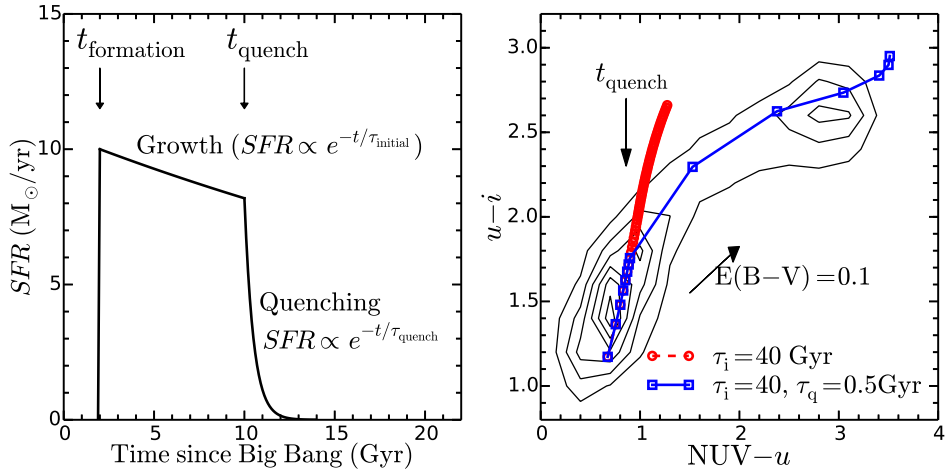}
\caption{Schematic plot to show the two-phase SFH adopted in stellar population model. 
The left panel directly shows the two-phase SFH definition and the four parameters that 
characterize the SFH. The right panel shows the model applied with the one-phase SFHs
(red circles) and two-phase SFH (blue squares)
in NUV$-u$ v.s. $u-i$ color-color diagram. The black contour represents the distribution of our galaxy sample.
The black arrow indicates the color changing caused by extinction with color excess E(B$-$V)=0.1.
}
\label{figure1}
\end{figure*}

The left panel of Figure 1 illustrates the two-phase SFH definition.
There are four parameters in the two-phase SFH: formation time $t_{\rm f}$, 
quenching start time $t_{\rm q}$, initial e-folding time
$\tau_{\rm i}$ for the star-forming stage and e-folding time $\tau_{\rm q}$ for the quenching stage. 
The right panel shows the one-phase (red circles) and two-phase (blue squares) models
in NUV$-u$ v.s. $u-i$ color-color diagram. The black contour represents the distribution of our galaxy sample.
It is worth noting that a clear turn of galaxy distribution is present at NUV$-u$ 
$\sim 1.4$ and
a dramatic number density drop can also be seen {in} between the blue cloud and the red sequence.
The symbols on the model track mark the evolution time from 1 Gyr to 14 Gyr with an interval of 1 Gyr.
For the one-phase model, we extend the end point to 60 Gyr to illustrate the model's evolution trend.
It can be seen that, the model track with two-phase SFH successfully reproduce the angled distribution 
and sparse region in between the main populations. 
However, the one-phase model, with the e-folding time given by \citet{noeske2007}, 
significantly deviate from the angled distribution of galaxies. Meanwhile, the number density profile of the one-phase model 
in NUV$-u$ color space is rather flat, conflicting with the clearly observed low density region.
{In other words}, the galaxy distribution in NUV$-u$ v.s. $u-i$ color-color diagram is a strong evidence 
for the two-phase evolution scenario.

For the star-forming galaxies, 
\citet{noeske2007} found that the cosmic evolution of main-sequence relation
can be well explained assuming a SFH of single exponential decline. They also derived  
the e-folding time and formation time as functions of stellar mass in Equation (6) of \citet{noeske2007}. 
Generally, more massive galaxies tend to be formed at earlier 
epoches and their SFR decline more rapidly with a shorter e-folding time.
For example, galaxies with masses of $10^{10.1} {\rm M_{\odot}}$, $10^{10.3} {\rm M_{\odot}}$, $10^{10.5} {\rm M_{\odot}}$
formed at redshift $z\sim1$, $1.5$, $1.8$ 
with e-folding times of 40, 25, 16 Gyr, respectively.
In later analysis, we do not try to fit the properties of 
galaxies one by one but only use the models to illustrate the statistical
evolutionary trend. 
 
We generate model SED using \citet{bc03} population model spectra with \citet{chabrier2003} initial
mass function. The metallicity is fixed to the solar value. 
Following \citet{scha2014} we ignore contribution by stellar population younger than 3 Myr because they tend to be surrounded by optically thick clouds
and are not visible in the UV and optical.
In addition, we apply dust extinction to the model SED with E(B$-$V)=0.1 assuming \citet{cardelli1989} extinction law.
This amount of extinction is broadly consistent with the estimates by stellar continuum fitting
\citep{ossy2011} for our galaxy sample on average.
Since young stellar populations are probably embedded in birth molecular cloud and suffer higher extinction than the 
old stellar population, we assume that the stellar population younger than 10 Myr suffer dust extinction three times higher than 
the general population. 
As quiescent galaxies shows negligible dust extinction, we apply a degressive dust extinction to the model after 
quenching which drops at a speed proportional to the SFR decline and reaches zero at NUV$-u$ of 3 mag.
Finally we obtain the model colors by convolving the model SED with {\em GALEX} and SDSS filter transmission curves.

\subsection{evolution speed in transition stage}
Currently, it is still not well understood
how fast SFR declines when a galaxy quenches or how fast galaxies evolve through the transition zone.
The evolution speed can be characterized by 
the e-folding time under the assumption of an exponential decline. 
Generally, the NUV$-u$ color traces the recent star 
formation activity and is sensitive to the fast SFR decline. On the other hand, the $u-i$ color is more sensitive 
to older stellar population and roughly indicates the average stellar age. Therefore, the slope of transition galaxies
in NUV$-u$ v.s. $u-i$ color will effectively constrain the e-folding time or evolution speed of quenching process. 
\citet{scha2014} provided broad constraint to the e-folding time of transition galaxies with different morphologies using NUV$-u$ v.s. {$u-i$} diagram. 
In this work, we investigate the e-folding time of quenching for galaxies splitted by stellar mass rather than morphologies.

Figure 2 shows the NUV$-u$ v.s. $u-i$ diagram for the galaxies in three mass bins from $10^{10}$ to $10^{10.6} {\rm M_{\odot}}$. 
The gray points represent the galaxies in each mass bin, while the contours denote the distribution of the whole data sample.
The black arrow indicates the effect caused by dust extinction with E(B$-$V)=0.1 Extinction by dust would introduce scatter 
to the galaxy distribution in the diagonal direction.
In each panel, lines are the toy models generated to illustrate the 
parameter range. 
The symbols on the model track represent the evolution time from 1 Gyr to 14 Gyr with an interval of 1 Gyr.
As mentioned above, we adopt model SFH of two exponential declines with 
e-folding times of $\tau_{\rm i}$ and $\tau_{\rm q}$. 
The legend in the bottom right shows the value of $\tau_{\rm i}$ and $\tau_{\rm q}$ for each model. 
We adopt the mass-dependent $\tau_{\rm i}$ given by \citet{noeske2007} because it provides a good 
description to the cosmic evolution of the star-forming main-sequence.
As we will argue below, galaxies fall in the green valley must have quenched recently, thus for simplicity, we 
assume their $t_{\rm q}$ to be the current time. For the three stellar mass bins, their $t_{\rm q}$ 
are 8/9/10 Gyr for the model in panels
from the left to the right, respectively.
It can be seen that, in all three mass bins, the model with $\tau_q$ of 0.5 Gyr best matches the
angled distribution 
compared to those models with $\tau_q$ of 1 and 0.2 Gyr.
Models with higher $\tau_q$ systematically offset from the data points with redder $u-i$ color, while 
others with lower $\tau_q$ deviate to the opposite direction. 
Therefore, statistically, the e-folding time of quenching process should have an average around 0.5 Gyr 
and rarely to be faster than 0.2 Gyr or longer than 1 Gyr.

Our results are broadly consistent with \citet{scha2014} who 
obtained a wide range of e-folding time for green valleys from 2.5 Gyr for late-type ones
to 0.25 Gyr for early-type ones.
The higher e-folding time derived by \citet{scha2014} can be due to their
different two-phase SFH which comprises a constant SFR (or 
infinite e-folding time) in the first 9 Gyr
and a subsequent exponential decline. This difference in SFH
result in a different starting point for quenching. As shown in Figure 7 in \citet{scha2014}, their
model start quenching at a point much bluer than the main population of star-forming galaxies.
This offset is more significant if we limit to galaxies with $M_*\sim 10^{10}{\rm M_{\odot}}$. 
Using the SFH given by \citet{noeske2007} provide a much more realistic starting point, matching the median 
color of star-forming galaxies of a given stellar mass. Thus this approach should provide a better description of the 
evolution speed during quenching.

\begin{figure*}
\includegraphics[width=\textwidth,viewport=150 0 1400 450,clip]{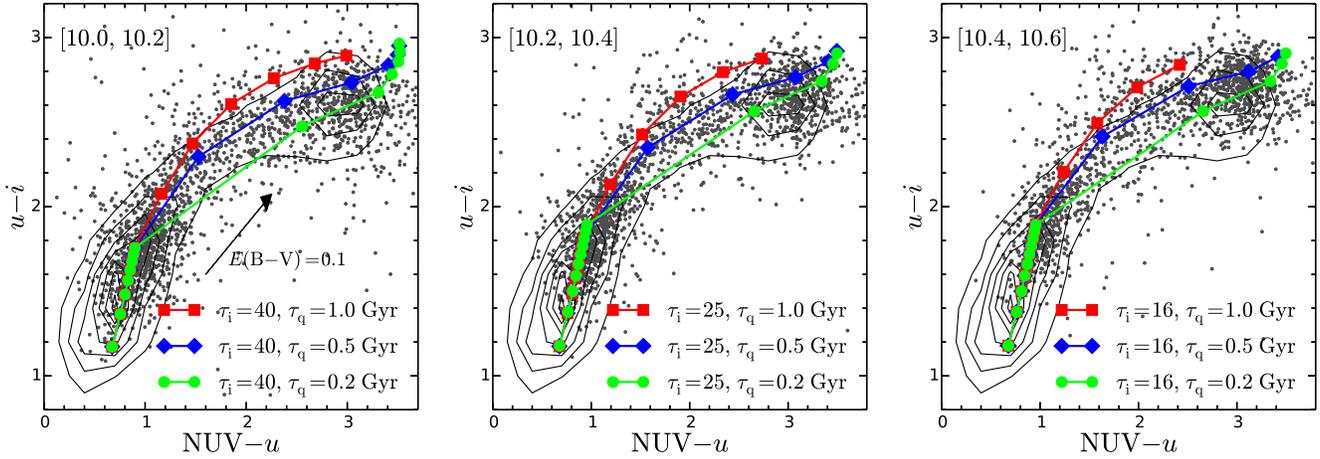}
\caption{NUV$-u$ versus $u-i$ color-color diagram. Black contours represent the distribution of the whole sample. 
The gray points are galaxies in each mass bin. Various lines represent the stellar population synthesis models as described in text. 
The legend in bottom right shows the e-folding time of star-forming and quenching stage for each model track. 
}
\label{figure2}
\end{figure*}

\section{Quenching fraction and quenching rate}
\subsection{number density profile}
So far, we have only used the information about the envelope of the galaxies' distribution in color-color 
space. The number density profile of galaxies in color space is another 
independent and critical piece of information we can utilize to investigate how 
likely galaxies quench and evolve through the transition zone.

The upper left panel of Figure 3 shows the galaxies in NUV$-u$ colors versus stellar mass diagram, while the 
other three panels are the number density profile in NUV$-u$ color space for galaxies in three mass bins.
In the mass-color diagram, a clear bimodal distribution is present.
Interestingly, galaxies in the blue cloud and red sequence are parallel to each other and separated by a wide low density region 
which we refer as the transition zone hereafter.
Blue circles and red squares represent the model tracks that best match the 
galaxy distribution in NUV$-u$ v.s. $u-i$ color-color diagram for the low- and high-mass bins, respectively.
As indicated by the model tracks, galaxy mass barely changes after quenching while their NUV$-u$ color 
increase dramatically.
This is the basis for our assumption that 
galaxies with the same mass but different color represent different stages of the same evolutionary sequence 
from star-forming, through transition, and to the red sequence.
As we discussed above, the spread of star-forming galaxies in NUV$-u$ color could 
probably be mainly due to the distribution of internal dust extinction and small variations in 
SFR around the long-term average value.

In this work, we focus on galaxies with stellar mass in 
three narrow bins, from $10^{10}$ to $10^{10.6} {\rm M_{\odot}}$. Their number density profile are shown in the 
upper right and bottom panels of Figure 3. 
It can be seen that the low density transition zone 
is clearly separated from the star-forming zone
by a dramatic number density drop at NUV$-u$ of 1.4$\sim$1.6. 
Interestingly, the number densities in the transition zone are roughly the same among the three mass bins with 
$\sim 100$ objects in each color bin of the width $\triangle C_{\rm bin}=0.2$ mag.
In the higher mass bin, there are relative more passive galaxies than star-forming galaxies, which indicate 
the quenching process in high mass systems either has been happening for a longer period
or they evolve faster than low mass galaxies.
According to \citet{noeske2007}, more massive galaxies formed earlier compared to low mass systems. It is natural to 
assume that quenching started earlier among them. 
In addition, we have shown above that galaxies of different masses have a similar e-folding time during quenching. 
Therefore, the different ratios between red sequence and blue cloud at different masses are probably due to different quenching start time.

\begin{figure*}
\centering
\includegraphics[width=15cm]{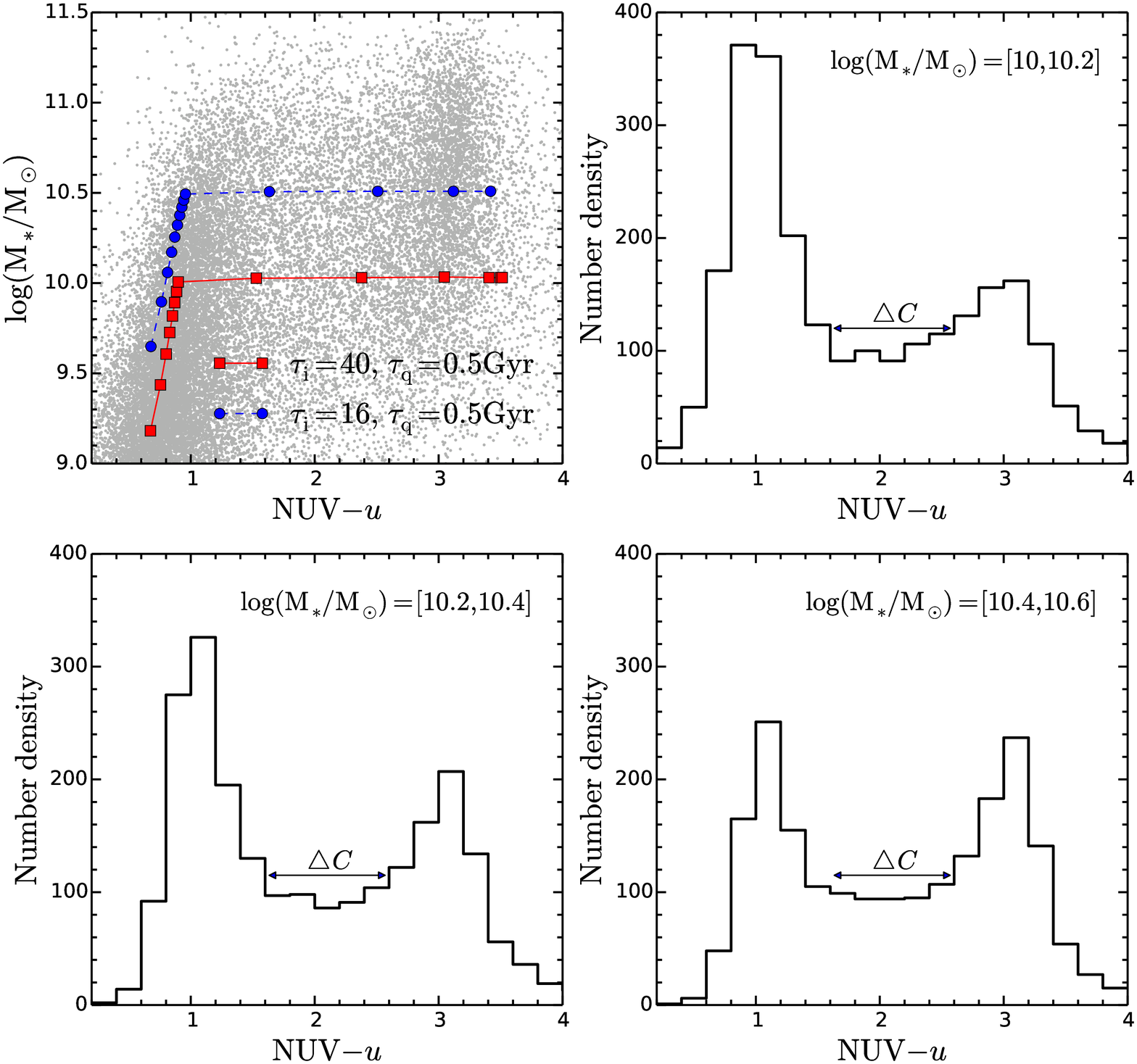}
\caption{Upper left panel: NUV$-u$ color versus stellar mass diagram.
Blue circles and red squares represent the model tracks which best represent the galaxy distribution in NUV$-u$ v.s.
$u-i$ color-color diagram in the low- and high-mass bin, respectively. 
Other three panels: Number density profile of galaxies in the each mass bin.
The color range of transition zone is marked with $\triangle C$.
}
\label{figure3}
\end{figure*}

\subsection{quenching rate}
The key to study the quenching fraction or quenching rate is to answer how the transition zone build up in the past.
First, we define the transition zone as the low density region between star-forming and passive galaxies.
As can be seen from NUV$-u$ v.s. mass diagram in Figure 3, the positions of the star-forming galaxies and red sequence 
galaxies are nearly constant in NUV$-u$, independent of stellar mass,
suggesting that the color range of transition zone should be similar for the 
galaxies in the three mass bins.
By visually inspecting the number density drop, we define a mass-independent transition zone with NUV$-u$ between 1.6 and 2.6 mag, 
which is marked in Figure 3 as $\triangle C$.
{We will verify this choice later when we fit the number density profile.}

\subsubsection{A simple approach}
We first adopt a simple approach to estimate the fraction of quenching. We assume all of the galaxies in transition zone 
are galaxies that quenched during a recent period $T$.
We will quantify the period $T$ in the next subsection. 
As a result, 503/476/489 galaxies are found in the transition zone from the low to high mass bins, respectively.
These similar total number in different mass bins are expected from the number density profile in Figure 3.
We define the quenching fraction to be the ratio of the transition galaxies to the sum of the transition galaxies 
and star-forming galaxies, because we expect the transition galaxies were part of the star-forming population at a time period T ago. 
This yields quenching fractions 
in the three mass bins to be 28\%, 31\%, and 40\%, respectively. 
We include these results in Table 1.
It should be noted these value of quenching fraction correspond to a time period of $T$.

\subsubsection{A more accurate approach}
The estimates of the simple method can be inaccurate in two aspects.
First, the red end of the transition zone may have contributions from galaxies that quenched more than $T$ Gyrs ago. 
Second, since there is a spread in color among star-forming galaxies, most recently quenched galaxies may still be having 
the same color as other slightly redder star-forming galaxies. 
For a more accurate estimate of quenching fraction, 
we simulate how the star-forming and transition zone build up.
It should be noted that there are several basic assumptions adopted in this approach as following:
\begin{enumerate}
\item At a given stellar mass, star-forming galaxies of different NUV$-u$ colors have the same probability
of being quenched.
\item The fraction of quenching is constant over the time interval $T$.
\item Quenched galaxies evolve through the NUV$-u$ color space at a constant speed.
\item Assumptions 1 and 3 above also apply to the time period between $T$ and 2$T$.
\end{enumerate}
  
We check the third assumption of constant evolution speed using the stellar population models. Figure 4 shows the 
evolution speed (i.e. color changing speed) of models in NUV$-u$ color space.
The dramatic increase {indicates} the time when quenching happens. 
It can be seen that the evolution speed of models is independent of the initial 
SFH ($\tau_{\rm i}$), and is roughly constant over the quenching phase.
The dashed red line 
represent the average evolution speed of the model from the beginning  
to the end of the transition stage. The variation of the evolution speed 
is less than 30\% and the assumption of a constant value is not far from truth.
In addition, the color changing speed in the last several Gyrs of the star-forming stage barely changes 
and is much smaller compared to that during the quenching stage.
 
Based on the assumption of a constant evolution speed, we could derive the crossing time $T$ of the transition zone.
For models with certain $\tau_i$ and $\tau_q$, we obtain the color-changing speed 
$v_i$ and $v_q$. 
Given that the color range of transition zone is caused by the differential evolution speed of the two evolution stages,
we have
\begin{equation}
T=\triangle C/(v_2-v_1)
\end{equation}
, where $\triangle C$ is the NUV$-u$ color range of the transition zone. 
As mentioned above, we 
adopt $\triangle C$ of 1.0 mag for the three mass bins as marked in Figure 3.
According to our results in \textsection3.2, the e-folding time of quenching stage is generally much 
shorter than that of the star-forming stage. Therefore we have $v_{\rm q} \gg v_{\rm i}$ and then 
the time interval $T$ should be determined by the e-folding time of the quenching stage.
If we adopt the best-fitting e-folding time of quenching stage of 0.5 Gyr, the time interval $T$ will be $\sim$1.5 Gyr. 

\begin{figure*}
\centering
\includegraphics[width=9cm]{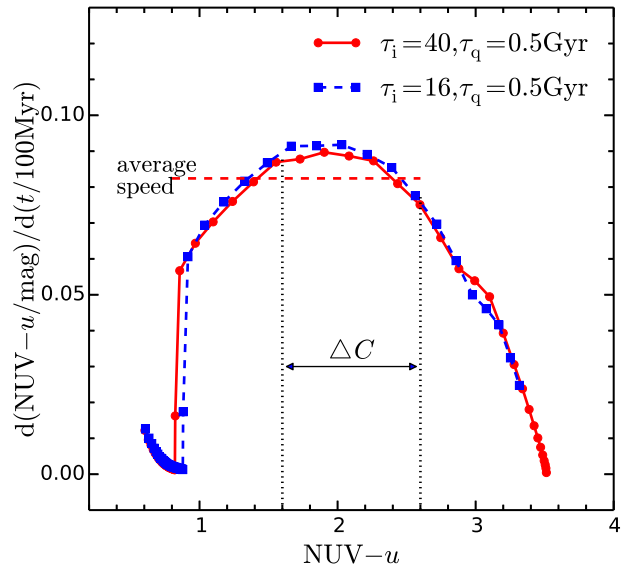}
\caption{Evolution speed of models in NUV$-u$ color space. The transition zone is marked as $\triangle C$.
The red dashed line denotes the average evolution speed of the model within quenching stage.
}
\label{figure4}
\end{figure*}

We run simulations for a number of quenching fractions ranging from 10\% to 60\%. For each trial quenching fraction,
we simulate the color distribution of the star-forming and transition galaxies based on the above assumptions and 
using partial information from the data. We consider galaxies in these color bins be made up of three components:
galaxies that have not yet quenched (star-forming), galaxies quenched 
during the last $T$ Gyr (`recently-quenched') and those that quenched more than $T$ Gyr ago (`older-quenched'). 
We obtain their numbers and distribution iteratively. At the end, we assess the overall matching of the color distribution using all
the color bins.

The simulation procedure can be described in five steps.
{\sl Step 1:} Initially, we assume the star-forming population $T$ Gyr ago is the sum of 
the galaxies in star-forming zone and transition zone today and it has the same NUV$-u$ distribution as the present star-forming 
population (given assumption 1 above). 

{\sl Step 2:} For a certain quenching fraction among these original star-forming galaxies,
we simulate the quenching process and
generate the distribution of the present 
star-forming population and `recently-quenched' population, based on the assumptions 2 and 3 above. 
Figure 5 illustrates the quenching simulation for the original star-forming galaxies in one NUV$-u$ color bin. 
Each red bar represents the galaxies quenched in each time interval, one fifth of $T$ Gyr. 
The number of `recently-quenched' galaxies declines slightly with time because the number of 
star-forming galaxies is declining, analogues to the radioactive decay of unstable isotopes.

{\sl Step 3:} Based on the distribution of `recently-quenched' population, we obtain the distribution 
of the `older-quenched' population. As implied by the definition of the `older-quenched' population, they are basically 
analogues of the present `recently-quenched' population in the star-forming zone which will evolve to be the `older-quenched'
population in the near future. Therefore, we apply the same distribution of the `recently-quenched' 
population in the star-forming zone to the `older-quenched' population, shifted in color by $\triangle C$ and nomalized 
so that the `older-quenched' population in the last color bin equal to the total number in that bin 
minus the number of `recently-quenched' population.

Now, we have generated a `older-quenched' population. We can refine the total number of the original star-forming population 
$T$ Gyr ago and its distribution. To update its distribution, we again follow today's distribution of star-forming population.
This needs to exclude `recently-quenched' population in the star-forming zone. Therefore, we use the data and the `older-quenched' population 
derived above to update the `recently-quenched' population. 

{\sl Step 4:} We update the 'recently-quenched' population in two parts: those in the star-forming zone and those in the transition zone. 
For the `recently-quenched' polulation in the transition zone ($1.6<$NUV$-u<2.6$), we take the difference between the data and 
the `older-quenched' population. For those in star-forming zone (NUV$-u<1.6$), we use the same distribution of previous version, 
but normalized 
to the new `recently-quenched' population in the first color bin of the transition zone.

{\sl Step 5:} The new distribution of the star-forming population is then set to be the observed distribution in star-forming zone
{after subtracting} the new distribution of `recently-quenched' population. 
With these new distribution of star-forming population and new number of `recently-quenched' population, we 
then rerun step 2-5 above until the total number of the star-forming population stops changing.

\begin{figure*}
\centering
\includegraphics[width=15cm]{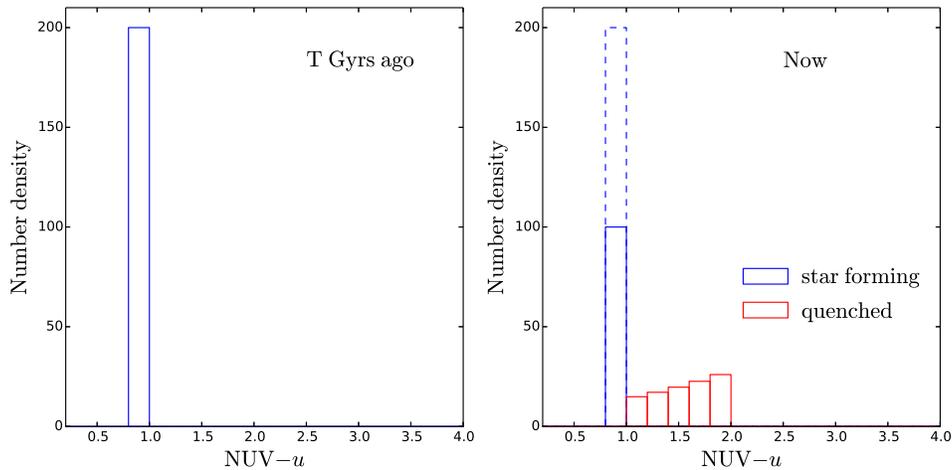}
\caption{Illustration for simulation of quenching process for original 
star-forming galaxies in one NUV$-u$ color bin from $T$ Gyrs before to present.
}
\label{figure5}
\end{figure*}

\begin{figure*}
\centering
\includegraphics[width=\textwidth,viewport=140 0 1400 450,clip]{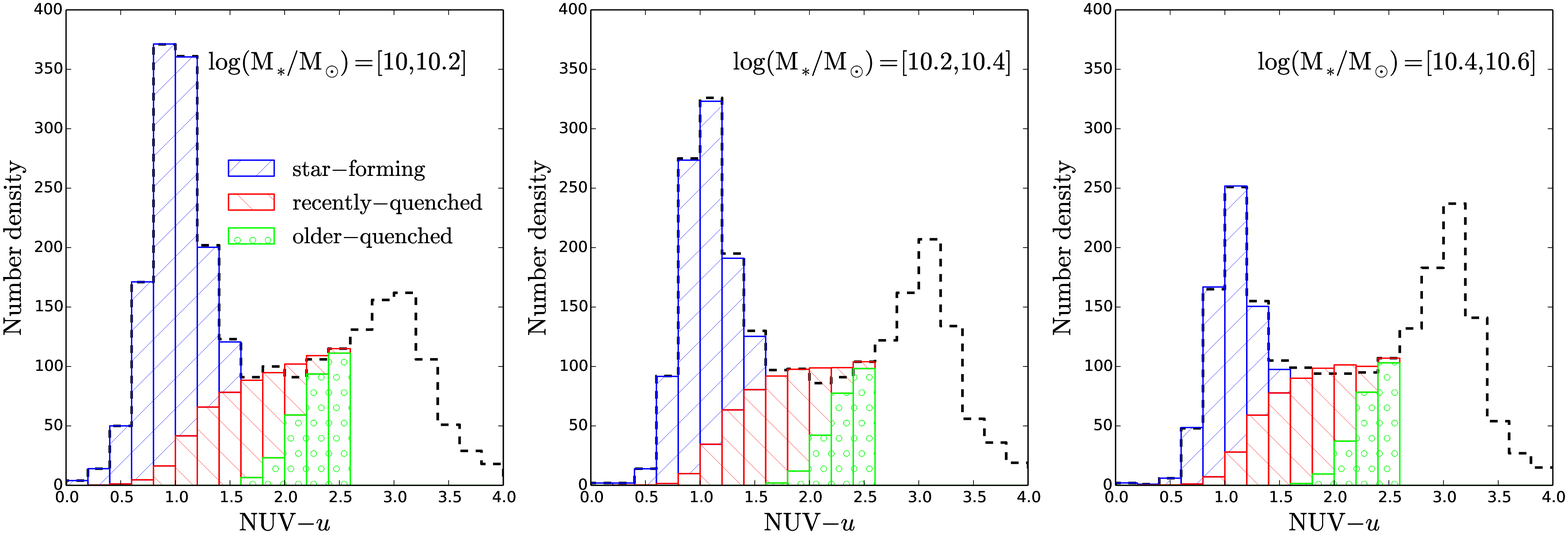}
\caption{Decomposition for number density profile of star-forming and transition galaxies. Blue bars represent the present star-forming 
galaxies and red bars represent galaxies quenched during the last $T$ Gyr. 
Green bars are the galaxies that quenched more than $T$ Gyr ago. The black solid bars denote the sum of these three components. 
We also show the number density profile of the data sample
as dashed histogram to guide the eye.}
\label{figure6}
\end{figure*}

Although part of the observed color distribution has been used in the iteration above, 
for each value of the quenching fraction, the simulated distribution (sum of the three components after Step 2) 
would not necessarily match the data in all the color bins. 
To assess how the simulated distribution match the data, we simply calculate {the average} absolute difference between 
the simulated distribution and the observed one {in each colour bin}.  
Finally, we obtain the best-fitting simulated distribution as shown in Figure 6 with 
quenching fractions of 28\%, 35\%, and 45\% for the three mass bins from low mass to high mass, respectively.
The results of the quenching fraction are summarized in Table 1. 
These quenching fractions are remarkably consistent with the 
simple estimates in \textsection4.1. 
Adopting an e-folding time of quenching stage of 0.5 Gyr which results in a $T$ of 1.5 Gyr, we could derive an quenching rate 
of 19\%/Gyr, 25\%/Gyr and 33\%/Gyr for the three mass bins, respectively.

{Considering the color range of the transition zone is chosen visually according to the number density drop, we also run the simulation above 
for other color ranges defining the transition zone. For each trial color range, we obtain a best-fit model color distribution.
The average difference in the number of galaxies per bin between model and data provides an indication of the fitting quality.
Changing the lower and upper limit of the color range by $\pm0.2$ result in much worse fitting, with the number difference per bin 
much larger than that expected given Poisson staticstics. The only change to the color range that still result in reasonable fit 
is to decrease the red limit by 0.2, i.e. defining the transition zone as [1.6,2.4].
In this case, the quenching fraction will be 23\%, 31\%, and 41\% for the three mass bins. Noting that the 
crossing time $T$ will also reduce to 1.2 Gyr. Therefore the quenching rate will be 19\%/Gyr, 27\%/Gyr, and 36\%/Gyr, only slightly higher
than the previous one. This confirms the 
robustness of our result.}
{However, considering} the large spread of galaxies around the best-fitting model in the NUV$-u$ v.s. $u-i$ color-color diagram in Figure 2, 
notable uncertainties 
are expected in the e-folding time of quenching stage and thus in the derived quenching rate.

\citet{moustakas2013} derived the quenching rate 
as a function of mass and redshift
by investigating the stellar mass function (SMF) evolution from $z\sim1$ to 0.
The obtained quenching rate increases with decreasing redshift.
However, the dependence of the rate on stellar mass is not monotonic with redshift. 
At $z\sim$ 0, the quenching rate tends to be lower in higher mass bins, opposite to our expectation.
However, at slightly higher redshift, this relationship reverse to higher quenching rate 
for higher mass bins.
For star-forming galaxies at low redshift with stellar mass range adopted in this work, 
the quenching rate obtained by \citet{moustakas2013} ranges between $\sim$7\%/Gyr$-$18\%/Gyr, which are
systematically lower than our results by a factor of 2 on average. Based on optical luminosity function, \citet{blanton2006} 
found that, from $z\sim1$ to 0 ($\sim$8 Gyrs), only $\sim$25\% star-forming galaxies must have quenched to match the pile-up of passive galaxies. 
This corresponds to a even lower quenching rate.

There could be several reasons that could account for the discrepancy in the quenching rate obtained in this work and that in literature. 
One possible contributor is the passive galaxies with rejuvenated star formation located in the 
transition zone.
These rejuvenated passive galaxies show low state recent or ongoing star formation over the global old stellar population \citep{fang2012}.
Recently, a significant fraction of passive galaxies in local universe
($\sim$10\%$-$30\%)
are found showing signatures of star formation in the UV \citep{yi2005,donas2007,schawinski2007} or IR \citep{shapiro2010}.  
The origin of these rejuvenated episodes is still under debate and several potential mechanisms are proposed, such as 
non-axisymmetric structures (e.g. bars), gas-rich mergers (e.g., \citealt{kaviraj2009}) and gas accretion from intergalactic medium.
Roughly speaking, a rejuvenating rate of 20\%/Gyr, which is typically found in local universe, can easily account for the 
discrepancy of quenching rate in this work and \citet{moustakas2013}.
Another reason for the different quenching rate can be the inaccuracy in the e-folding time of quenching stage.
If we adopt a higher e-folding time of 1 Gyr for the quenching stage, the crossing time $T$ 
will double and thus the quenching rate will shrink by a factor of 2. 
However, in the NUV$-u$ v.s. $u-i$ color-color diagram, the model with e-folding time of 1 Gyr does not represent the 
majority of galaxies in the transition zone. 
Some potential inaccuracies in stellar population model and/or dust extinction could possibly induce 
a higher e-folding time at the quenching phase that is comparable with color-color diagram. 
We will investigate in detail the reason for the difference in quenching rate 
in future work.

Until now, the mechanisms for triggering quenching is still not well understood. 
Physical processes that shutdown the cold gas supply or speed up the gas depletion rate are needed to cease the star formation in galaxies.
Several candidates are proposed, such as 
strangulation due to shock heating in massive halos,  
AGN feedback both in `quasar-mode' triggered by major mergers and `radio-mode' in massive galaxies
, bars that drive cold gas into the central region of galaxies
where it will be consumed quickly, and ram-pressure stripping in dense regions (see a detailed discussion in 
\citealt{scha2014} and reference therein).
Our results of quenching rate should provide important constraints on the proposed quenching processes and  
their implementation into galaxy evolution models.

\begin{table*}
  \caption{Quenching fraction for galaxies of different masses {(derived for transition zone defined as 1.6$<$NUV$-u<$2.6)}.}
  \label{table:1}
  \begin{tabular}{cccc}
\hline\hline
 & mass bin1 & mass bin 2 & mass bin 3 \\
 & [10.0, 10.2] & [10.2, 10.4] & [10.4, 10.6] \\
\hline
Simple estimation & 28\%  & $31$\% & $40$\% \\ 
Accurate estimation & 28\% & 35\% & 45\% \\
\hline
\end{tabular}\\
\end{table*}

\section{Summary}
In this work, we investigated the evolution speed of galaxies in the quenching stage and the quenching fraction 
of star-forming galaxies. 
We select a local galaxy sample from SDSS with {redshift} in [0.02, 0.05] 
and target a sub-sample of them with stellar mass in [$10^{10}$, $10^{10.6}$]${\rm M_{\odot}}$.
To reproduce their UV$-$optical colors we use stellar population synthesis models.
For simplicity, we adopt a model SFH of two-phase exponential decline which corresponds to 
the secular star-forming stage and the rapid quenching stage. 

In the NUV$-u$ v.s. $u-i$ color-color diagram, we find a clear turn along with a dramatic number density drop
which strongly support the two-phase evolution scenario.
To constrain the e-folding time of quenching stage,
we compare the model tracks with the transition galaxy distribution in NUV$-u$ v.s. $u-i$ color-color diagram. 
For the quenching stage, the e-folding time should be statistically within [0.2, 1] Gyr while
an e-folding time of 0.5 Gyr best matches the observed galaxy distribution.
Adopting this best-fitting e-folding time, the crossing time $T$ of the transition zone in NUV$-u$ color
is about 1.5 Gyr.

The galaxy number density profile in NUV$-u$ color space provide critical insights into the quenching fraction
during the past period $T$. Using a simple approach and a complicated but more accurate approach we have found 
consistent results for the quenching fraction among star-forming galaxies. 
In the three mass bins, the quenching fraction are found to be 28\%/35\%/45\%.
Adopting an e-folding time of 0.5 Gyr for the quenching stage which corresponds to a 
crossing time $T$ of 1.5 Gyr, we further derive an quenching rate of 
19\%/Gyr, 25\%/Gyr, and 33\%/Gyr for star-forming galaxies in the three mass bins.
{We also examine different definitions 
of transition zone and found our result is insensitive to slight changes of the transition zone definition.}

Compared to the quenching rate derived by \citet{moustakas2013} based on stellar mass function of star-forming and passive galaxies
from $z\sim1$ to 0, our results are broadly consistent within uncertainties but systematically higher. This 
discrepancy, to our knowledge, is possibly due to the passive galaxies with rejuvenated star formation 
in transition zone and potential inaccuracies in derived 
e-folding time at the quenching stage. Therefore our results could be upper limits of quenching rate.

\section*{Acknowledgements}
We thank Michael Blanton, Jeremy Tinker, and ChangHoon Hahn for helpful conversations about this work. 
J. Lian gratefully acknowledge support from China Scholarship Council. 
This work is supported by the Strategic Priority Research Program "The Emergence of Cosmological 
Structures" of the Chinese Academy of Sciences (No. XDB09000000), the National Basic Research Program of 
China (973 Program)(2015CB857004), and the National Natural Science Foundation of China (NSFC, Nos. 11225315, 1320101002, 11433005 and 11421303).

{\sl Galaxy Evolution Explorer (GALEX)} is a NASA Small Explorer, 
launched in 2003 April. We gratefully acknowledge NASA’s support for construction, operation and science analysis for the
{\sl GALEX} mission, developed in cooperationwith theCentre National d'EtudesSpatiales of France and the Korean Ministry of Science and Technology. 

Funding for the SDSS and SDSS-II has been provided by the Alfred P. Sloan Foundation, the Participating Institutions, the National Science Foundation, 
the U.S. Department of Energy, the National Aeronautics and Space Administration, the Japanese Monbukagakusho, the Max Planck Society,
and the Higher Education Funding Council for England. The SDSS Web Site is http://www.sdss.org/.

The SDSS is managed by the Astrophysical Research Consortium for the Participating Institutions. 
The Participating Institutions are the American Museum of Natural History, Astrophysical Institute Potsdam, 
University of Basel, University of Cambridge, Case Western Reserve University, University of Chicago, 
Drexel University, Fermilab, the Institute for Advanced Study, the Japan Participation Group, Johns Hopkins University, 
the Joint Institute for Nuclear Astrophysics, the Kavli Institute for Particle Astrophysics and Cosmology, the Korean Scientist Group, 
the Chinese Academy of Sciences (LAMOST), Los Alamos National Laboratory, the Max-Planck-Institute for Astronomy (MPIA), 
the Max-Planck-Institute for Astrophysics (MPA), New Mexico State University, Ohio State University, University of Pittsburgh,
University of Portsmouth, Princeton University, the United States Naval Observatory, and the University of Washington.

\end{document}